\documentclass[prb,preprint]{revtex4}

\usepackage{amssymb}
\usepackage{graphicx}
\usepackage{amsmath}
\usepackage{bm}

\newcommand{\half}{\frac{1}{2}}

\newcommand{\eps}{\epsilon}
\newcommand{\dd}{\partial}

\newcommand{\E}{ {\bf E}}
\newcommand{\B}{ {\bf B}}
\newcommand{\A}{ {\bf A}}
\newcommand{\jj}{ {\bf j}}
\newcommand{\J}{ {\bf J}}
\begin{document}

\title{Time (in)dependence in general relativity}
\author{S. Deser}
\email{deser@brandeis.edu}
\affiliation{Department of Physics, Brandeis University, Waltham,
Massachusetts 02454}
\affiliation{Lauritsen Laboratory, California Institute of Technology,
Pasadena, California 91125}

\author{J. Franklin}
\email{jfrankli@reed.edu}
\affiliation{Department of Physics, Reed College, Portland, Oregon 97202}

\begin{abstract}
We clarify the conditions for Birkhoff's theorem, that
is, time-independence in general relativity. We work primarily at the
linearized level where guidance from electrodynamics is particularly useful.
As a bonus, we also derive the equivalence principle. The basic
time-independent solutions due to Schwarzschild and Kerr provide concrete
illustrations of the theorem. Only familiarity with Maxwell's equations and tensor analysis is required.
\end{abstract}

\maketitle

\section{Introduction}
A major obstacle to teaching general relativity is the initially
confusing mathematics underlying useful, physical simplifications. We focus
in this paper on the conditions that lead to the simplest regime,
time-independence. Because general relativity is
coordinate-invariant, what does it mean to speak of a particular
coordinate's independence? The answer is illuminating. Loosely, we
expect that there exists a choice of coordinate frame in which the
gravitational field does not depend on $t$. But is this a
meaningful, that is, invariant criterion? The answer is yes: it means that
the spacetime geometry allows the existence of a Killing vector field
$f_{\mu}(x)$ that obeys the tensor equation 
\begin{equation}
D_\nu f_\mu + D_\mu f_\nu \equiv \dd_\nu f_{\mu} + \dd_\mu f_{\nu} -
g^{\sigma\rho} (\dd_\nu g_{\mu\rho} + \dd_\mu g_{\nu\rho} - \dd_\rho
g_{\mu\nu} ) f_\sigma = 0,
\label{du}
\end{equation}
where $g_{\mu\nu}$ is the metric and $D_\mu$ is the covariant derivative
with respect to it, as defined in Eq.~\eqref{du}. We use the signature
$(-+++)$ and units such that $c=1$. If $f_\mu$ is also timelike ($f^2 <
0$), then the solution in the frame where $f_\mu = g_{0\mu}$ (more
manifestly, the contravariant form $f^\mu$ of the vector is $f^\mu =
\delta^\mu_0$) implies that
\begin{equation}
\label{Kframe}
\dd_0 g_{\mu\nu} = 0,
\end{equation}
and there is no time dependence. (A special property of time-independent
geometries is that in (and only in) them, matter systems such as particles
retain a conserved energy, just as in flat space.)

Our main point is that we have re-expressed the issue of when a
given geometry is time-independent, that is, when there exists a frame where
Eq.~\eqref{Kframe} holds, as a covariant (coordinate-independent) criterion:
the existence of solutions to Eq.~\eqref{du}. All this transcription makes no reference to field equations. There exist many frames
where
$t$-dependence is present, but that is not the point. It is not true false
that every geometry has a static frame -- the Killing equation is a strong
requirement.

\section{Maxwell} 
We begin with electrodynamics whose field equations outside sources, unlike
general relativity, can be written entirely in terms of gauge invariant
field strengths,
\begin{subequations}
\label{VacMax}
\begin{align}
\nabla \cdot \E = 0\\
\nabla \cdot \B = 0 \\
\dot \E = - \nabla
\times \B \\
\dot \B = \nabla \times \E.
\end{align}
\end{subequations}
The $\dot \E$ equation's longitudinal part (see the following) implies that
$\dot
\E^L = 0$, which exhibits the fact that the ``Coulomb" part of $\E$ is always
time-independent, whatever the behavior of the interior charges. The
remaining, dynamical transverse part $\E^T$ and its partner $\B$
(transverse by definition) cannot depend on time if they vanish
identically, which is the case for spherically symmetric configurations: any
$\E(r)$ is necessarily of the form $\nabla S(r)$ and is purely longitudinal.
There is no monopole radiation; it is also the only guaranteed static case,
as dipole and higher configurations define transverse vectors.
Equation~\eqref{VacMax} does not therefore require time-dependence,
or electro/magneto-statics would not exist.

For future use we recall that the transverse/longitudinal division of any
vector field ${\bf V}$ is a decomposition of unity, 
\begin{equation}\label{Vpartition}
V_i = \big[ (\delta_{ij} - \hat k_i \hat k_j) + \hat k_i \hat k_j \big] V_j,
\end{equation}
along some arbitrary unit vector direction $\hat k$. Its more familiar Fourier
transform is
\begin{equation}
{\bf V}= {\bf V}^T + {\bf V}^L,
\end{equation}
where $\nabla \cdot {\bf V}^T = \nabla \times {\bf V}^L = 0$.

Our discussion has been couched in terms of the gauge invariant field
strengths $\E$ and $\B$, whose time (in-)dependence is unaffected by the
choice of gauge. The underlying potentials $(A_0, \A)$ are another story:
even if
$(\E,\B)$ are static, there exist gauge choices for which the potentials do
depend on $t$ by adding gauge terms $\dd_\mu \Lambda(r,t)$ that do not
affect $F_{\mu\nu} =\dd_\mu A_\nu-\dd_\nu A_\mu$. In any case the transverse
vector potentials are unaffected, being gauge invariant. Only 
$(A_0,\A^L)$ can be altered, keeping $\E^L$ unchanged.

It is instructive to analyze the equations in terms of the $A_\mu$ in
parallel with the general relativity discussion in Sec.~\ref{sec:gr} where
potentials are unavoidable.

\section{general relativity}\label{sec:gr}
For our purposes the gravitational field is a glorified tensor version of
the vector Maxwell field $A_\mu$, and we expect similar properties of the results there to apply. At the linearized level, the Einstein equations outside
sources are
\begin{equation}
\label{LinG}
2 G_{\mu\nu} \equiv \Box h_{\mu\nu} - \big(\dd_\mu \dd_\alpha h^\alpha_{\
\nu} + \dd_\nu \dd_\alpha h^\alpha_{\ \mu} \big) + \dd_\mu \dd_\nu h -
\eta_{\mu\nu}
\big(\Box h - \dd_\alpha \dd_\beta h^{\alpha\beta}\big) = 0
\end{equation}
for the field $h_{\mu\nu}$ with $h \equiv h^\alpha_{\ \alpha}$; all
indices are moved by the Minkowski metric $\eta_{\mu\nu}$. As for Maxwell's
equations, we decompose Eq.~\eqref{LinG} into space and time
components, with the simplifying notation $h_{0i} \equiv N_i$ and $h_{00}
\equiv N$. The theory is invariant under linearized gauge/coordinate
transformations
$h_{\mu\nu}
\rightarrow h_{\mu\nu} + \dd_\mu \xi_\nu + \dd_\nu \xi_\mu$, that is,
$G_{\mu\nu}(\dd_\mu \xi_\nu + \dd_\nu \xi_\mu)=0$, an invariance that is
useful to exploit.

The component form (the linearized version of a
decomposition used long ago to analyze the full theory\cite{ADM}) of
Eq.~\eqref{LinG} is
\begin{subequations}
\label{eq:set}
\begin{align}
2 G_{00} &= \nabla^2 \tilde h - \dd_i \dd_j h_{ij} \\
2 G_{0i} &= \nabla^2 N_i - \dd_j \dot h_{ji} - \dd_i \dd_j N_j + \dd_i
\dot{\tilde h} \label{Graw}\\
2 G_{ij} &=\Box h_{ij} + \dd_i \dot N_j +
\dd_j \dot N_i - (\dd_i \dd_k h_{kj} + \dd_j \dd_k h_{ki} ) \nonumber \\
& \quad +
\big(\delta_{ij} \nabla^2 - \dd_i \dd_j\big) (N - \tilde h) + \ddot{\tilde
h} \delta_{ij} + \delta_{ij} \big(\dd_m \dd_n h_{mn} - 2 \dd_k \dot
N_k\big), 
\end{align}
\end{subequations}
with $\tilde h \equiv h^i_{\ i}$ the trace of the spatial part of the
field. This slightly complicated set of equations simplifies when we
decompose the spatial tensors $h_{ij}$ and the vectors $N_i$, the latter into
transverse/longitudinal parts via Eq.~\eqref{Vpartition}, the former by the
following partition of unity:
\begin{subequations}
\begin{align}
\label{hpartition}
h_{ij} &= h^{TT}_{ij} + h^T_{ij} + \dd_i h_j + \dd_j h_i, \\
\dd_i
h^{TT}_{ij} & = \dd_i h^{T}_{ij} = 0 = h^{TT}_{ii} \\ h_{ij}^T &= \half \big(
\delta_{ij} - \nabla^{-2} \dd_i \dd_j \big) h^T.
\end{align}
\end{subequations}
The six components of $h_{ij}$ are decomposed linearly, orthogonally,
and uniquely into two TT (transverse traceless), one T (traceless), and three
$h_i$ parts. Any spatial tensor equation thus consists of three independent
sets. The four quantities $(h_i ,N^L_i)$ are pure gauges (variables that can be
arbitrarily changed by using the gauge freedom of the theory) that cry out
to be set to zero, leaving the gauge invariant set $(h_{ij}^{TT}, h^T,
N_i^T, N)$ once we use the available gauge invariance. Now
Eq.~\eqref{eq:set} reduces to
\begin{subequations}
\label{Gfixball}
\begin{align}
2G_{00} &= \nabla^2 h^T = 0 \label{Gfixa}\\
2G_{0i} &= \nabla^2 N_i^T + \dd_i \dot h^T = 0\label{Gfixb} \\
2G_{ij} &= \Box h_{ij}^{TT} + \big(\dd_i \dot N_j^T + \dd_j \dot N_i^T\big)
+
\big(\delta_{ij} \nabla^2 - \dd_i \dd_j\big) \big(N - \half h^T\big)
\label{Gfixc}\nonumber \\
& \quad+ \half (\delta_{ij} + \nabla^{-2} \dd_i
\dd_j) \ddot h^T = 0 .
\end{align}
\end{subequations}

The
time-independence of $h^T$ follows from the longitudinal part of
Eq.~\eqref{Gfixb}, and the relation $N = \half h^T$ follows from
Eq.~\eqref{Gfixc}. This seemingly innocuous equality is none other than the
expression of Einstein's principle of equivalence. This expression of the equivalence principle even applies to full GR.\cite{ADM} The latter states that (in suitable
units) the inertial and gravitational masses of every physical system are
equal. Inertial mass/energy is the conserved quantity that (in the linear
regime) sums over the $T_{00}$ contributions of the interior sources. This sum is the monopole moment of the Poisson equation \eqref{Gfixa}
(if we restore
$T_{00}$ as its right-hand side); hence it is the coefficient of the
leading $1/r$ term in $h^T$. In contrast, gravitational mass is a
very different quantity that determines the system's gravitational pull,
the ``Newtonian" force, on slow particles. (Einstein implicitly assumed the
existence of static frames, as we have also established here.) This force is
the gradient of the leading $1/r$ part of $h_{00}$. Thus, in
general relativity the field equation \eqref{Gfixc} enforces the
universal equality of the desired $1/r$ coefficients.

The time-independence
of $N^T_i$ results from the transverse vector part of Eq.~\eqref{Gfixc}: The
four ``Newtonian" components of the field are time-independent outside
sources. Time dependence can reside only in the remaining $\Box
h_{ij}^{TT}$ dynamical modes, namely those field components unaffected by the choice of
gauge and undetermined by the interior sources. Hence $t$-independence is
forced whenever TT tensors are forbidden. Spherical symmetry is one such
case, because all spherically symmetric tensors have the form
\begin{equation}
S_{ij}(r) = \delta_{ij} A(r) + \dd_i \dd_j B(r),
\end{equation}
and so, by Eq.~\eqref{hpartition} have no TT parts. This result is the
basis of the Birkhoff theorem:~\cite{BHOFF} all spherically symmetric
configurations are also time-independent, a result valid also in full
general relativity.

Unlike Maxwell, there is another category of
fields lacking a TT part, namely those with dipole character. As we saw
there, dipoles permit a transverse vector, but their single direction is
not generic enough to construct a TT tensor. Axial symmetry
does permit TT, for example via the tensor harmonic $P_2(\cos \theta)$. To
summarize at this point, both Maxwell and linearized general relativity
gauge fields only allow time-dependence of their true dynamical excitations,
and only when those modes can be present, which always excludes
spherical symmetry and also dipole symmetry for the general relativity case.

\section{Kerr and Schwarzschild}
It is instructive, at the linearized level, to relate the exterior solution
properties to explicit matter sources. In electrodynamics the current
consists of two parts: the charge density $\rho$ and the longitudinal
current $\jj^L$, which obey the continuity equation $\dot \rho + \nabla \cdot
\jj^L = 0$, and the transverse current $\jj^T$. The $(\rho,\jj^L)$
subset couples only to the longitudinal electric field, which is 
equivalent to it, and as we saw, is time-independent away from sources. The
transverse electric and magnetic fields are generated by the transverse
current and can be time dependent if $\jj^T$ is. Similar reasoning
applies to general relativity: the source here is the tensor $T_{\mu\nu}$,
whose
$(T_{00}, T_{0i}^L)$ components are like $(\rho, \jj^L)$. They obey the same
continuity equation and excite only the metric component $h^T$, which is
also $t$-independent outside of source distributions. Because general
relativity is a tensor theory, there is another ``charge" associated with
momentum like
$T_{00}$ was with energy, namely $(T_{0i}, T_{ij}^L)$, which also obeys
continuity and is coupled to $N_i^T$. The remaining source part,
$T_{ij}^{TT}$, which may, but need not, depend on time, excites the dynamical $h_{ij}^{TT}$ fields. 

An important
example of time-independence is furnished by the Kerr
solution~\cite{Kerr,BL} of full general relativity, which we will reproduce
in the following. In our linearized context, the static metric is generated
by a time-independent spinning point mass with
\begin{equation}
\label{SpinSource}
T_{00}= m \delta^3({\bf r}), \quad T_{0i}= a m \eps_{ijk} s_j \dd_k
\delta^3({\bf r}),
\end{equation}
where $s_j$ denotes the (constant) unit spin vector. 
As explained in Ref.~\onlinecite{MTW} the space integral of $T_{00}$ is 
the total mass $m$, and that of $T_{0i}$ vanishes because there is no momentum.
Its first moment, the angular momentum $\J$, is given by $\J= a m
{\bf s}$. The notation choice that expresses $J\sim am$ is historical, but
has the virtue that $m=0$ is actually just flat space (also in full general
relativity) and the parameter $a$ reduces to that defining ellipsoidal
coordinates in ordinary euclidean 3-space. The opposite limit, $a=0$,
defines the spherically symmetric static Schwarzschild
solution.

We will not discuss in detail the full general relativity
extensions of our linear results. Consider, without deriving it (there is
no simple way to do so) the full Kerr interval
\begin{equation}\label{Kerr}
ds^2 = -g_{tt} dt^2 + g_{rr} dr^2 + g_{\theta\theta} d\theta^2 + g_{\phi \phi} d\phi^2 + 2 g_{t \phi} dt d\phi.
\end{equation}
There are five functions of $(r,\theta)$ which are (in units of $c = 1 = 16
\pi G$),
\begin{subequations}
\label{Kerrfunctions}
\begin{align}
g_{tt} &= -(1 - 2 M r/\rho^2)\\
g_{rr} &= \rho^2/\Delta\\
g_{\theta\theta}
&= \rho^2 \\
g_{\phi\phi} &= \sin^2\theta \big[r^2 + a^2) + 2 a^2 M r
\sin^2\theta/\rho^2\big] \\
g_{t\phi} &= -2 a M r \sin^2\theta/\rho^2,
\end{align}
\end{subequations}
with
$\rho^2 \equiv r^2 + a^2 \cos^2\theta$ and $\Delta \equiv a^2 - 2 M r +
r^2$. 

The linearized limit of Eqs.~\eqref{Kerr} and \eqref{Kerrfunctions}, or equivalently its asymptotic
form, is a superposition of the (linearized) Schwarzschild solution and a spin term 
$h_{0 \phi}$ corresponding to the source \eqref{SpinSource}
\begin{subequations}
\label{LinKerr}
\begin{align}
h_{00} &= \frac{2 m}{r}\\
h_{0\phi} &= -\frac{2 a m \sin^2\theta}{r} \\
h_{ij} &= \frac{2 m}{r} \frac{x_i
x_j}{r^2}.
\end{align}
\end{subequations}
We emphasize that the time-independence here is derivable directly from the
exterior equations, apart from details of the interior source, as
we would expect for a spinning spherical ball of charge in E\&M, its natural
analogue.\cite{FB}

\section{Conclusions}
By working primarily in the linearized limit, we have provided, using the
Maxwell template, a framework for understanding the basis of
time-independence in general
relativity in terms of the underlying physics and source geometry. Our main
conclusion is that the time-dependence of solutions of gauge theories such as
Maxwell's or general relativity is a property of their radiation modes. If
these are forbidden due to spherical (dipole) symmetry, then
time-independence is guaranteed. In particular, the Kerr and Schwarzschild
solutions illustrate the absence of dipole and monopole
excitations. Although the full general relativity is unavoidably more
complicated (and involves global issues we have bypassed here), our results
capture at least its long distance properties.

\begin{acknowledgements}
We are grateful to Prof.\ J.\ Hartle for stimulating criticism that led to
this (we hope) improved version of our earlier paper. This work was
supported by NSF grant PHY-04-01667.
\end{acknowledgements}


\begin{thebibliography}{5}

\bibitem{ADM} R. Arnowitt, S. Deser, and C. W. Misner, ``The dynamics of
general relativity," in \textsl{Gravitation: An Introduction to Current
Research}, edited by L. Witten (John Wiley \& Sons, New York, 1962).
Reprinted as gr-qc/0405109.

\bibitem{BHOFF} For the history and a modern derivation, see S. Deser and J.
Franklin, ``Schwarzschild and Birkhoff a la Weyl," Am. J. Phys. {\bf 73}
(3), 261--264 (2005), gr-qc/0408067.

\bibitem{Kerr} Roy P. Kerr, ``Gravitational field of a spinning mass as an
example of algebraically special metrics," Phys. Rev. Lett. {\bf 11} (5),
237--238 (1963).

\bibitem{BL} Robert H. Boyer and Richard W. Lindquist, ``Maximal analytic
extension of the Kerr metric," J. Math. Phys. {\bf 8} (2), 265--281 (1967).

\bibitem{MTW} Charles W. Misner, Kip S. Thorne, and John Archibald Wheeler,
\textsl{Gravitation} (W. H. Freeman, New York, 1973).

\bibitem{FB} J. Franklin and P. T. Baker, ``Linearized Kerr and spinning massive bodies: An electrodynamics analogy," Am. J. Phys., to appear.

\end{thebibliography}
\end{document}